\def\year{2023}\relax
\documentclass[letterpaper]{article} % DO NOT CHANGE THIS
\usepackage{aaai22}  % DO NOT CHANGE THIS
\usepackage{times}  % DO NOT CHANGE THIS
\usepackage{helvet} % DO NOT CHANGE THIS
\usepackage{courier}  % DO NOT CHANGE THIS
\usepackage[hyphens]{url}  % DO NOT CHANGE THIS
\usepackage{graphicx} % DO NOT CHANGE THIS
\usepackage{natbib}  % DO NOT CHANGE THIS AND DO NOT ADD ANY OPTIONS TO IT
\usepackage{caption} % DO NOT CHANGE THIS AND DO NOT ADD ANY OPTIONS TO IT
\usepackage{booktabs}
\usepackage{multicol}
\usepackage{amsmath}
\usepackage{enumitem}
\usepackage{multirow}
\usepackage{todonotes}
\usepackage{pifont}% http://ctan.org/pkg/pifont
\title{What are Your Pronouns? Examining Gender Pronoun Usage on Twitter}

\author{Julie Jiang\textsuperscript{\rm 1, \rm 2}, Emily Chen\textsuperscript{\rm 1, \rm 2}, Luca Luceri\textsuperscript{\rm 1}, Goran Murić\textsuperscript{\rm 1}, Francesco Pierri\textsuperscript{\rm 3}, \\Ho-Chun Herbert Chang\textsuperscript{\rm 4}, Emilio Ferrara\textsuperscript{\rm 1, \rm 2, \rm 5}}
\affiliations{% Affiliations
    \textsuperscript{\rm 1} Information Sciences Institute, University of Southern California, Marina Del Rey, CA, USA \\
    \textsuperscript{\rm 2} Department of Computer Science, Viterbi School of Engineering, University of Southern California, Los Angeles, CA, USA\\
    \textsuperscript{\rm 3} Politecnico di Milano, Italy \\
    \textsuperscript{\rm 4} Department of Quantitative Social Science, Dartmouth College, Hanover, NH, USA\\
    \textsuperscript{\rm 5} Annenberg School of Communication, University of Southern California, Los Angeles, CA, USA \\ 
    juliej@isi.edu, echen@isi.edu, lluceri@isi.edu, gmuric@isi.edu, francesco.pierri@polimi.it, \\
    herbert@dartmouth.edu, ferrarae@isi.edu}
\begin{document}

%%


%   <concept>
%       <concept_id>10003120.10003121.10011748</concept_id>
%       <concept_desc>Human-centered computing~Empirical studies in HCI</concept_desc>
%       <concept_significance>500</concept_significance>
%       </concept>
%   <concept>
%       <concept_id>10003120.10003130.10011762</concept_id>
%       <concept_desc>Human-centered computing~Empirical studies in collaborative and social computing</concept_desc>
%       <concept_significance>500</concept_significance>
%       </concept>
%     <concept>
%         <concept_id>10010405.10010455.10010461</concept_id>
%         <concept_desc>Applied computing~Sociology</concept_desc>
%         <concept_significance>300</concept_significance>
%     </concept>
%  </ccs2012>

% \ccsdesc[500]{Human-centered computing~Empirical studies in HCI}
% \ccsdesc[500]{Human-centered computing~Empirical studies in collaborative and social computing}
% \ccsdesc[300]{Applied computing~Sociology}



\maketitle
\begin{abstract}
The increasing awareness of nonconforming gender identities puts discussions of gender inclusivity at the forefront. Using people's preferred pronouns is a primary way to develop gender-inclusive language. This work presents the first empirical research on the self-disclosure of gender pronouns on social media. Leveraging a Twitter dataset with over 2 billion tweets collected over two years, we find that the public self-disclosure of gender pronouns is on the rise. The disclosure of gender pronouns is particularly popular among users employing \textit{she} series pronouns, followed by \textit{he} series pronouns, and a smaller but sizable amount of non-binary pronouns. By analyzing Twitter messages and users' sharing activities, we identify the users who choose to disclose their gender pronouns and additionally distinguish users of various gender pronoun categories. Analyzing the relationship between social network exposure to gender pronouns and gender pronoun adoption, we show that those who frequently interact with users who disclose gender pronouns are also more likely to adopt gender pronouns in the future. This work carries implications for research on gender inclusivity.
\end{abstract}
\section{Introduction}
Discussions and disclosures of gender pronouns are becoming ubiquitous at schools, workplaces, and especially on social media \cite{waterloo,wsj2021why,reuters2020they}. Online platforms such as Twitter, Instagram, and LinkedIn are rolling out designated fields for users to add their preferred pronouns in their biographies \cite{reuters2021why}. Many proactively ask, ``What are your pronouns?'', a question that symbolizes ``an invitation to declare, to honor, or to reject, not just a pronoun, but a gender identity'' \cite{baron2020what}. Encouraging gender pronoun sharing fosters gender-inclusive language, which is especially important for people whose genders do not fit neatly into the stereotypical binary gender roles. Non-binary gender pronouns are constantly evolving as gender identity awareness grows \cite{wp2019guide,gustafsson2021pronouns}. The Swedish gender-inclusive, third-person pronoun \textit{hen} achieved widespread usage and increasing acceptance since its inception in the Swedish vocabulary in the early 2010s \cite{gustafsson2021pronouns}. In English, the singular \textit{they} is the most widely accepted gender-neutral pronoun \cite{hekanaho2020generic}, followed by popular picks such as \textit{xe} and \textit{ze} \cite{them2020gender}.
% \footnote{\url{https://en.wikipedia.org/wiki/Gender_neutrality_in_languages_with_gendered_third-person_pronouns\#Table_of_standard_and_non-standard_third-person_singular_pronouns}}.
Preferred pronouns can also shift for those who are gender fluid, moving across the spectrum of gender identities over time \cite{harvard2020gender,diamond2020gender}.
\begin{figure}
  \fbox{\includegraphics[width=0.95\linewidth]{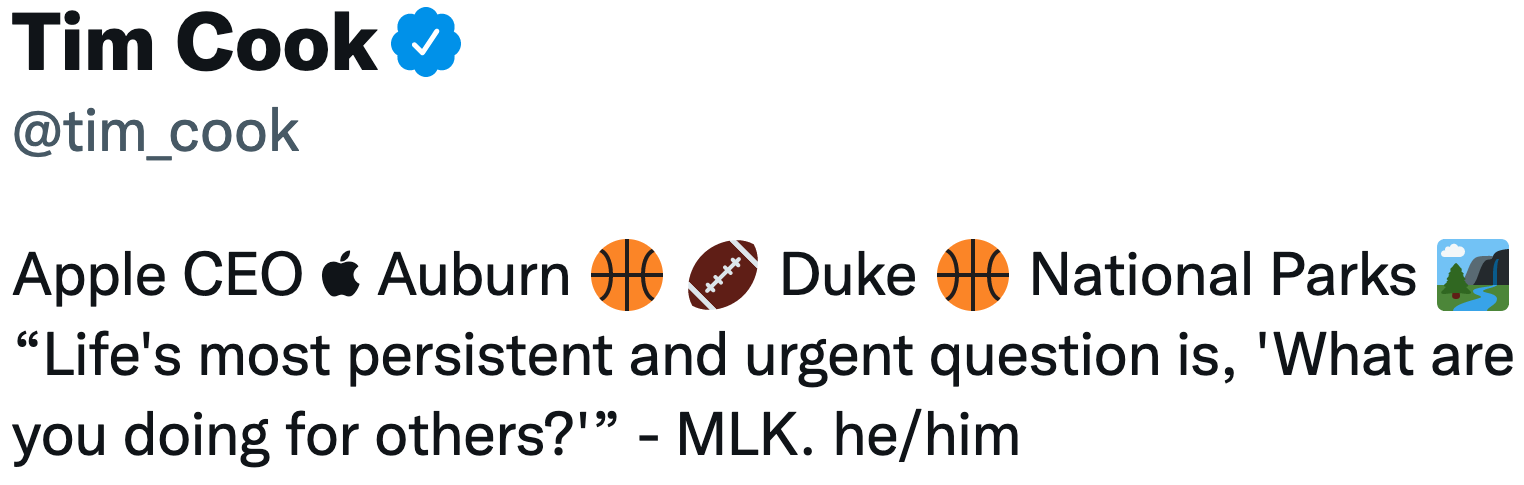}}
  \caption{The Twitter biography of the public figure Tim Cook shows he uses the \textit{he} series pronouns (Jan 2023).}
  \label{fig:timcook}
\end{figure}
Acceptance of minority gender identities faces many challenges \cite{pew2017views,pew2017republicans,pew2022deep,pew2022americans}, with 54\% of the current US population believing that gender identity is determined by physiological attributes \cite{pew2022americans,pew2022attitudes}. However, the share of the population who publicly use non-binary pronouns or state themselves as gender-nonconforming is undeniably rising \cite{pew2019about,pew2022about}: the proportion of people who personally know someone who uses gender-neutral pronouns grew from 18\% in 2017 to 26\% in 2021 \cite{pew2019about}. In 2020, an estimated 5\% of US young adults identify as gender-nonconforming \cite{pew2022about}.

In light of the increasing awareness of gender nonconformity, this paper offers an empirical and analytical approach to provide an initial understanding of gender pronoun disclosure on social media. In particular,  we seek to answer the following research questions:
\begin{itemize}
    \item \textbf{RQ1:} \textit{What is the prevalence of gender pronoun usage in Twitter user biographies over time?}
    We show that though the proportion of users who use gender pronouns remains a minority at less than 8\% at any given time, this number is rising over time, up 33\% from 2020 to 2021. Twice as many tweets are posted by \textit{she} series (6M) compared to those shared by \textit{he} series users (3M). A notable 1.6M tweets are posted by users with non-binary pronouns.
    \item  \textbf{RQ2:} \textit{Can we automatically identify the presence and type of gender pronouns in a user's biography?} 
    We can detect a user's gender pronoun category, or lack thereof, from their Twitter usage and sharing activities.
    \item  \textbf{RQ3:} \textit{Can we automatically identify gender pronoun adoption from prior social network exposure?} 
    We show that users who adopted gender pronouns, especially non-binary pronouns, experienced higher social network exposure to gender pronouns than those who did not.
\end{itemize}

To address these RQs, we leverage a large recent longitudinal Twitter dataset on COVID-19 collected over two years (2020-2021) \cite{chen2020tracking}. As discussions on COVID-19 are not primarily focused on gender disclosure, this dataset grants us access to a large sample of users who may or may not use gender pronouns. We also acknowledge that similar conclusions were found in a parallel work \cite{tucker2022pronoun}, strengthening the validity of our findings. Our findings provide actionable insights toward a principled understanding of gender identity and gender inclusivity on social media and stimulate further research.
\section{Background}

\subsection{Gender Identity and Pronouns}

\textit{Gender} is a deeply rooted societal concept that many believe to be determined by \textit{sex} at birth \cite{pew2022americans,pew2022attitudes}. But while \textit{sex} is a biological state, \textit{gender} is an identity that can be \textit{either} male or female, \textit{neither} male nor female, somewhere in between, or something else altogether \cite{marwick2013gender}. Gender identity also need not be immutable or singular \cite{keyes2018misgendering,keyes2021you}. The common social understanding of gender maps directly from physiology and dictates how people of each sex should behave, conventionally landing either in masculinity or femininity \cite{deaux1985sex,marwick2013gender,goffman1977arrangement,wolf2000emotional,witt2010self}. The existence of other arrangements of gender threatens this notion of well-defined boundaries of sex and gender \cite{stryker2013transgender}, defying gender norms deeply reinforced from popular culture to everyday discourse \cite{gauntlett2008media,cameron1998gender}. As such, gender nonconforming people face discrimination and stigma. One of the most devastating social phenomena is the high rates of suicide and depression among the marginalized LGBTQ+ community \cite{haas2010suicide,rotondi2012depression,budge2013anxiety,toomey2018transgender}, aggravated by targeted hate crimes and systematic violence against gender-nonconforming people \cite{hrc2020epidemic,cnn2020florida,gyamerah2021experiences}. Recognizing non-binary gender pronouns is a crucial step towards gender equity and inclusivity \cite{convo2021what,brown2020makes,knutson2019recommended,moser2019gender,parks2016gender}. Using a person's correct pronouns shows respect and validation for their chosen gender identity and self-representation \cite{parks2016gender,knutson2019recommended}. In this work, we utilize users' self-disclosed gender pronouns as a proxy for expressions of gender identity.

\subsection{Gender Pronouns on Social Media}
Most empirical gender research on social media treats gender as a binary variable \cite{ciot2013gender,alowibdi2013language,liu2013whats,bamman2014gender,zheng2016profile,fink2012inferring}. This is harmful in and of itself for the gender-nonconforming people we address in this work, perpetuating biases rooted in misgendering based on sexism and cisgenderism \cite{ansara2014methodologies,keyes2018misgendering}. A 2021 study by \citet{fosch2021little} found that Twitter's automatic, binary gender classifier misgendered 19\% of the people surveyed. Other studies focus exclusively on trans or other LGBTQ+ people \cite{krueger2015twitter,karami2018characterizing,tuah2020twitter}, neglecting the wider cisgender society and those who have not openly declared their gender. 

There are two gender-related studies on Twitter COVID-19 discourse. \citet{thelwall2021male} conducted a thematic analysis of the Twitter biographies of male, female, and non-binary people who tweeted about COVID-19 in the UK, demonstrating variations in their interests, jobs, relationships, sexuality, and linguistic styles. \citet{al2021investigating} explored the differences in gendered COVID-19 discourse co-occurring with men, women, and non-binary keywords, suggesting that women disproportionately experienced domestic violence and that non-binary people had concerns about blood donation. Building on past literature, we carry out a large-scale analysis of not only users who disclosed gender identity but also those who did not, as well as to analyze their Twitter activities beyond biography descriptions and text. 

A recent concurrent preprint by \citet{tucker2022pronoun} replicates several of our key findings. Instead of a focused COVID-19 dataset, they collected a longitudinal Twitter dataset of US users using the 1\% streaming API. They also find that the use of gender pronouns is on the rise and that there are substantially more \textit{she} series users. Further, users with gender pronouns tend to cluster together in the following network. In this work, by focusing on users who appear several times throughout the dataset, we further contribute by showing how the social network is linked to the adoption of gender pronouns over time.
\begin{figure*}
    \centering
    \includegraphics[width=\linewidth]{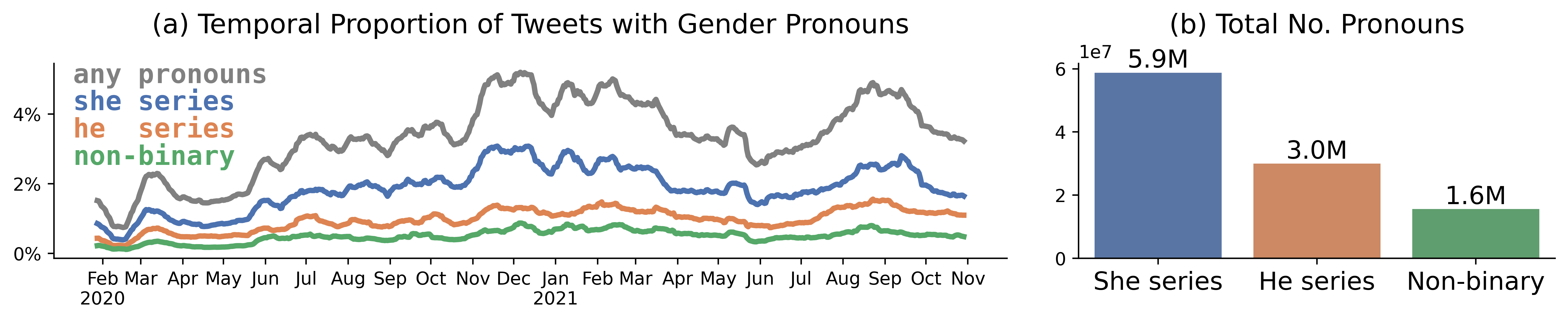}
    \caption{(a) Daily proportions of tweets (15-day rolling average) by GP users, displaying a statistically significant rising trend in gender pronoun usage (Mann-Kendall Trend test, $p<0.001$). (b) The total number of tweets by GP users.}

    \label{fig:at_a_glance}
\end{figure*}

\section{Data and Definitions}
We use a public collection of nearly two years of Twitter data by \citet{chen2020tracking} (release v2.72)\footnote{\url{https://github.com/echen102/COVID-19-TweetIDs}} from Jan 21, 2020, to Nov 5, 2021. The dataset was collected using keywords related to the COVID-19 pandemic and consists of over 2 billion tweets. It is important to note that the data was collected in real-time; therefore we can track changes in a user's gender pronoun status in their biography over time.

\subsection{Definitions}
To maintain consistency throughout this paper, we define the following terms and phrases. Each user is categorized either as a \textbf{GP} (gender pronoun) user who disclosed any gender pronouns or a \textbf{NP} (no pronoun) user who did not disclose gender pronouns. Importantly, we infer a user's gender from their disclosure of pronouns---we consider users to be \textbf{women} if they only use \textit{she} series pronouns and \textbf{men} if they only use \textit{he} series pronouns---but we caveat the gender and gender pronouns are not necessarily equivalent. Someone who uses \textit{she} series pronouns can be a cis-gendered woman, trans, non-binary, or gender-diverse person. Further, within the context of this paper, we use the umbrella term \textbf{non-binary} to refer to all those who use gender-nonconforming pronouns (e.g., \textit{they)}, gender expansive pronouns  (e.g., \textit{he/she}), or gender fluid pronouns (transitions from using \textit{she} to \textit{they}, \textit{he} to \textit{she}, \textit{they} to \textit{he}, etc.). Though we have no intentions of diminishing the complex and multiplicitous arrangements of gender identity \cite{keyes2021you}, we acknowledge the limitations of our rigid categorization. The rationale behind this choice is to initiate a high-level empirical analysis that can shed light on explicit self-disclosures of gender on social media at scale.

\subsection{Gender Pronoun Categorization}\label{sec:categorization}
We build a regular expression to match gender pronouns that appeared in users' biography descriptions or display names. We do not match gender pronouns from the tweet texts because pronouns are parts of speech frequently used in normal discourse. To be matched as a GP user, the user's biography must contain a substring with more than one gender pronoun separated by either forward slashes or commas, with or without additional blank spaces. We limit our scope to three categories of gender pronouns, which are the \textit{he} series pronouns: \{`he', `him', `his'\}, \textit{she} series pronouns: \{`she', `her', `hers'\}, and non-binary pronouns: \{`they', `them', `theirs', `their', `xe', `xem', `ze', `zem'\}. We further match the substring `any pronouns' as an additional indication of gender nonconformity. The non-binary pronouns are chosen due to their relatively widespread usage in society and in previous literature \cite{hekanaho2020generic}.  
% If only \textit{he} series pronouns or only \textit{she} series pronouns are detected, then we label that user with the gender they identify with. If they used non-binary pronouns, or a mix of gender pronouns from more than one category (e.g., `he/she', `she/they'), or have transitioned from using one category of gender pronouns to another, then we label them as non-binary.

\section{Gender Pronouns on Twitter}\label{sec:rq1}
% \begin{figure*}
%     \centering
%     \includegraphics[width=0.9\linewidth]{figures/top_gender_nop_desc_unigrams.png}
%     \caption{Top tokens that appeared in the biography of users who use the \textit{he} series pronouns, \textit{she} series pronouns, and non-binary pronouns, as well as users who do not have gender pronouns, ranked by weighted log-odds.}
%     \label{fig:desc_tokens}
% \end{figure*}
% \subsection{Overall Trends}

To address RQ1, we begin by looking at the overall gender pronoun usage on Twitter over time. Fig. \ref{fig:at_a_glance}(a) charts the temporal patterns of the daily proportion of tweets by GP users over almost two years. Tweets by GP users represent a minority of all users' tweets (mean = 3.3\%, median = 3.3\%, maximum = 7.3\%). That said, we observe a rising proportion of tweets by GP users. Using a Mann-Kendall trend test, we find this growth to be statistically significant $(p<0.001)$. In particular, the average proportion of tweets by GP users had a 33\% increase, growing from 2.86\% in 2020 to 3.82\% in 2021 ($t$-test, $p<0.001$).  We also see a consistent pattern of \textit{she} series users (5.9M) dominating the GP user pool, followed by \textit{he} series users (3M) and then a smaller but considerable amount of non-binary pronouns users (1.6M), as illustrated in Fig.~\ref{fig:at_a_glance}(b). The fact that there are twice as many tweets by \textit{she} series users as by \textit{he} series users stands in stark contrast with data collected via third-party Twitter surveys, which showed that Twitter users consist of 50\% women and 50\% men \cite{pew2019sizing}.  Given that is it not likely to have twice as many women as men in this Twitter dataset, our work suggests that women are more prone to declare their gender pronouns than men or that women tweet more often about COVID-19.

% \subsection{Biography Descriptors}\label{sec:bio_descriptors}
% Since we infer pronoun statuses and genders from biography descriptions, we investigate whether they embed other lexical signals tied with gender identity. To this end, we collect the frequencies of all the tokens (except for the pronoun tokens) used in the biographies of \textit{she} series, \textit{he} series, non-binary, and NP users. We rank the tokens using the weighted log-odds method with uniform Dirichlet priors \cite{monroe2008fightin}, comparing each set of tokens with the combined rest of the sets of tokens.

% Fig. \ref{fig:desc_tokens} shows the top tokens that appear in users' biographies. LGBTQ+ keywords are among the top tokens in GP users' biographies, with `gay' appearing in the biographies of \textit{he} series users and `queer', `nonbinary', `trans', `lesbian', `bi', and `nb' (short for `non-binary') appearing in those of non-binary pronouns users. Both \textit{she} and \textit{he} series users frequently use political keywords such as `blacklivesmatter' (`blm' for short) and `feminist' and use more interpersonal descriptors such as `mom', `wife', `auntie', `dad', and `husband'.  Non-binary pronoun users also use terms that suggest disability (`disabled', `autistic'). Our results echo those of \citet{thelwall2021male}, which showed that GP users frequently include descriptions of interpersonal relationships and reference politics. In contrast, NP users favor generic Twitter terms such as `news', `breaking', and `updates'.
% , as well as conservative politics-related keywords. 

\section{Detecting Gender Pronouns}
\label{sec:rq2}
In this section, we test if we can identify a user's gender pronoun status from their tweets, Twitter account metadata, and sharing activities. This analysis will help us understand the characteristics separating these users. 
% We first aim at identifying users who use \textit{any} gender pronouns, followed by a more fine-grained experiment aimed at distinguishing users of various gender pronoun categories.

\subsection{Method}
\subsubsection{Feature Sets.} We use the full span of the collected data and consider only users who had at least 10 posts to ensure we have sufficient data points per user. For each user, we build a series of feature sets as follows:
\begin{itemize}
\item \textbf{{Text}} ($\text{dim}=768 \times 20$): the BERTweet embeddings \cite{bertweet} of their most recent 20 tweets, zero-padded if they had less than 20 tweets. We use this value since most (60\%) of the users had less than 20 tweets. BERTweet is the state-of-the-art language model developed specifically for Twitter, producing 768-dimensional embeddings per tweet.

\item  \textbf{{Text Pronouns}} ($\text{dim}=7$): the number of times \textit{I}, \textit{you}, \textit{we}, \textit{he}, \textit{she}, \textit{it}, \textit{they}, and other gender-nonconforming pronouns appeared in their tweets (see Appendix).

\item  \textbf{{User}} ($\text{dim}=6$): The last entry of user-level metadata including whether they are verified and their number of friends, followers, favorites, statuses, and lists.

\item  \textbf{{Activities}} ($\text{dim}=9$): The number of original tweets, retweets, quoted tweets, and replies the user posted as well as the number of hashtags, URLs, mentions, characters, and tokens used. % we do not include individualism/collectivism here because they're already in Text Pronouns

\item  \textbf{{Mentions}} ($\text{dim}=\text{1K}$): We gather the top 1K most frequently mentioned users in our dataset and build a vector representing how frequently they were mentioned. Note that we treat retweets as a subset of mentions. 
\end{itemize}
We purposely omit user biography as a feature because it is our source of ground truth labeling of gender pronouns.
% and may contain obvious gender cues \cite{tucker2022pronoun}. 

\subsubsection{Modeling.}\label{sec:rq2_modeling}
We run two experiments. Experiment (1) is the binary classification task of detecting whether a user discloses gender pronouns or not; experiment (2) is the multiclass classification task of detecting the category of gender pronoun disclosed. To address data imbalance, we sample 10K users from each of the gender pronoun groups (\textit{he} series, \textit{she} series, and non-binary) followed by 30K NP users, totaling 60K users. We gather two such random samples of 60K users, one for hyperparameter-tuning and one for final testing via 5-fold CV. For experiment (2), the 30K NP users are not included. As we will show below, this decision is in part driven by the relative ease in which the models are able to distinguish users \textit{with} from users \textit{without} pronouns in experiment (1). In this way, we can investigate the factors that separate users of each gender pronoun group. We train and test a deep neural network (DNN) for each prediction task, using each feature set in isolation as ablation studies, as well as all feature sets in combination. For every input scenario, we fine-tune the model using randomized grid search to test over 40 combinations of hyperparameters, fixing 20\% of the data for validation. The exact model architecture and hyperparameters searched can be found in the Appendix. We use the macro-F1\footnote{Since the data is balanced, macro-F1 is equivalent to micro-F1.} score as the classification metric and use two baseline methods for comparison: \texttt{Random}, which predicts random labels based on the distribution of labels in the training set, and \texttt{Majority}, which predicts the majority label. 
\begin{figure}
    \centering
    \includegraphics[width=\linewidth]{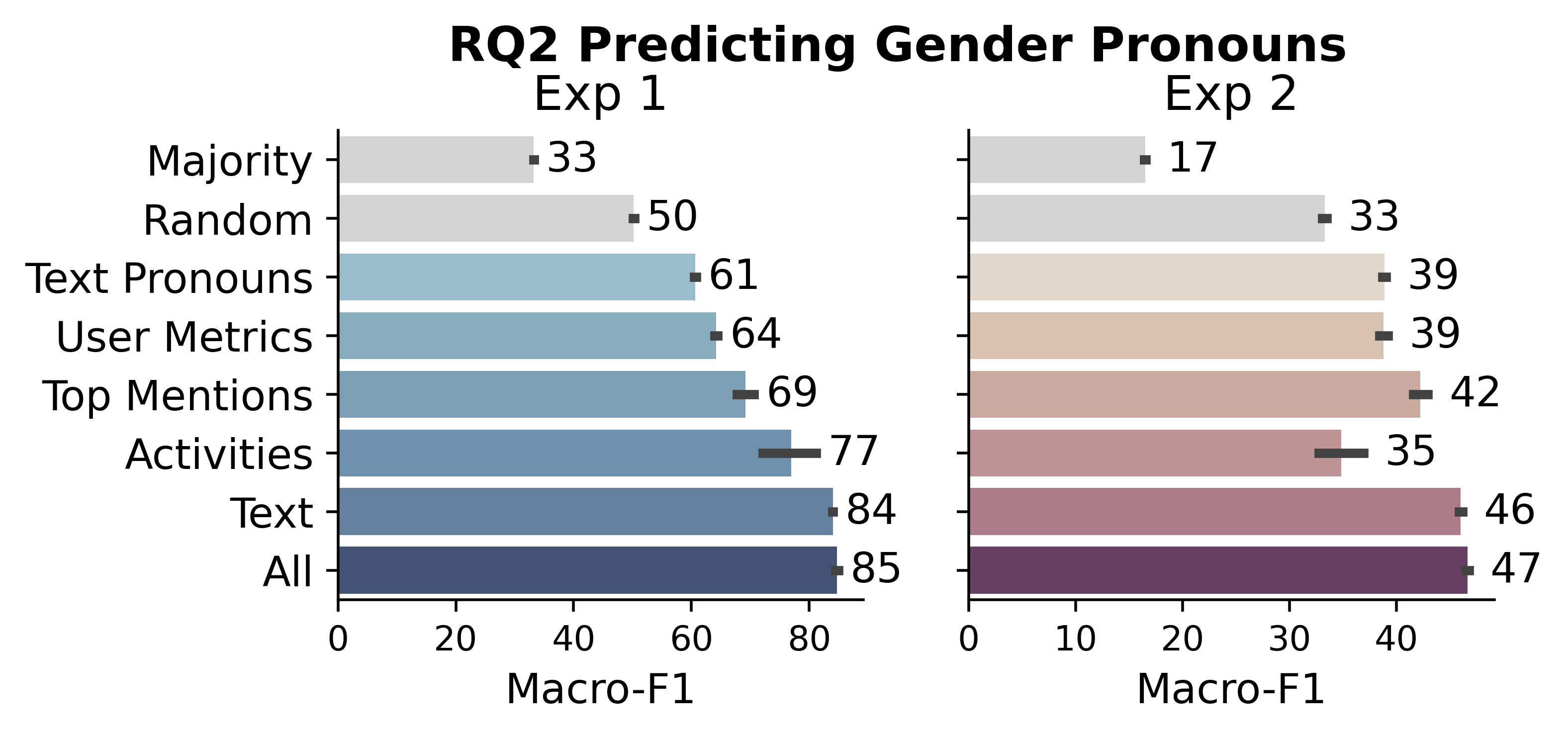}
    \caption{RQ2: Classification performance of the binary classification of identifying the presence of gender pronouns (Exp 1) and in the multiclass classification of the three gender pronoun types (Exp 2). Error bars reflect standard errors.}
    \label{fig:rq2_results}
\end{figure}
\begin{figure}
    \centering
    \includegraphics[width=\linewidth]{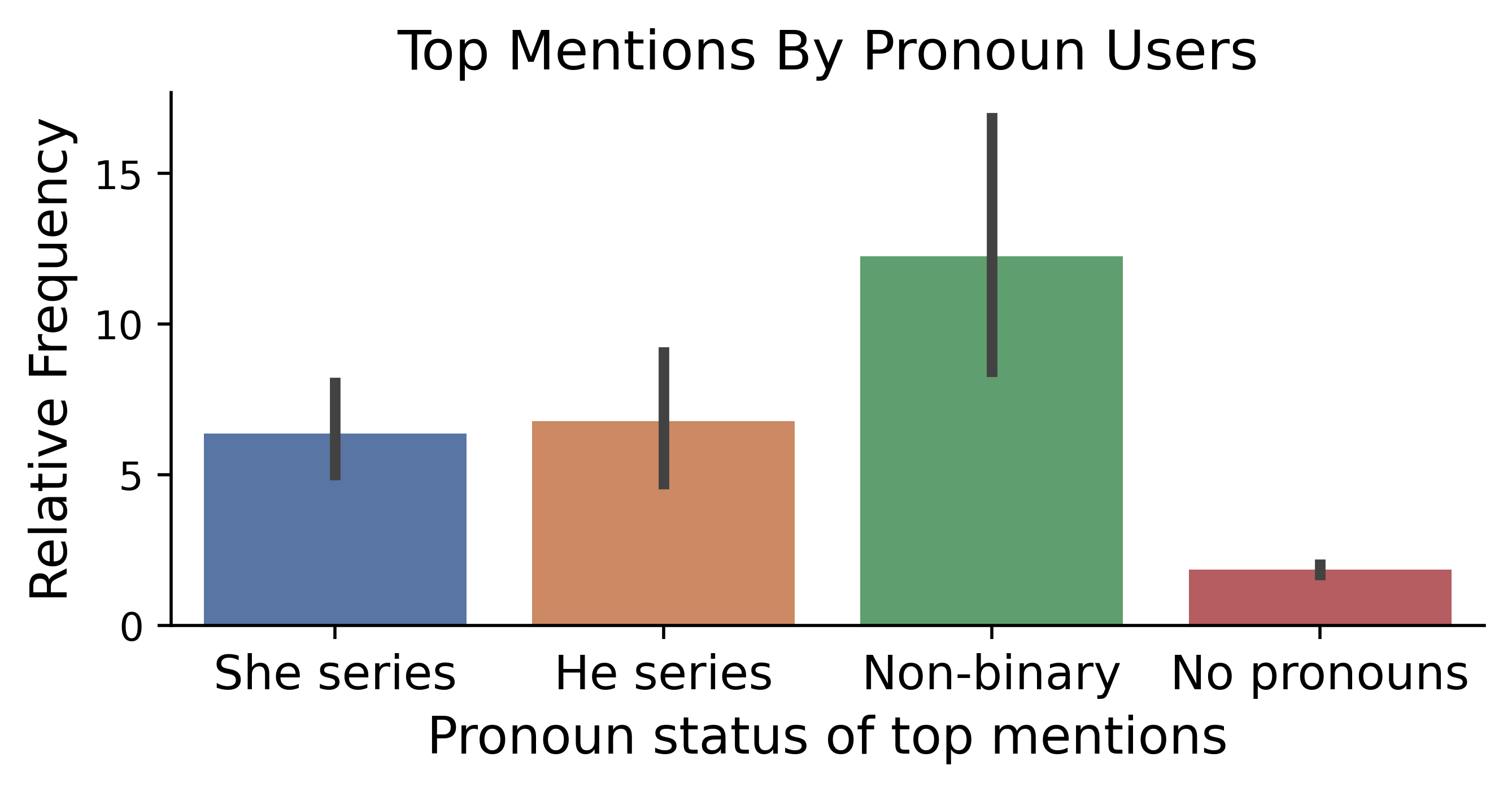}
    \caption{The relative frequency of a user being mentioned by a GP user compared to an NP user, limited to the top 1K highly mentioned users by all users. Users with pronouns, especially non-binary pronouns, are statistically significantly more likely to be mentioned by GP users.}
    \label{fig:rq2_mentions}
\end{figure}
\subsection{Results}

\subsubsection{Classification Performance.} For experiment (1), the binary classification distinguishing GP users from NP users, we find that using all of the feature sets improves predictive performance over the baseline \texttt{Random} and \texttt{Majority} models (Fig. \ref{fig:rq2_results}). Embeddings of the textual data are the best single predictors of gender pronouns, achieving a macro-F1 of 84\% ($SE=0.05\%$). User activities are the second best predictors at 77\% macro-F1 ($SE=4.53\%$), albeit with a much larger variance. The feature set on top mentions achieves an impressive score of 69\% macro-F1 ($SE=2.12\%$). Including all features yields a slightly better score of 85\% ($SE=0.24\%$). These results indicate that users' sharing activities encode substantial signals to predict whether they disclose gender pronouns. 
% We note that this high classification accuracy could be due to the differences between organizational and personal accounts, in which the former normally would not use gender pronouns.

For experiment (2), the multiclass classification of gender pronoun category, we note a similar improvement in classification performance over the baselines (Fig. \ref{fig:rq2_results}). However, the improvement is more modest than what was achieved in experiment (1), The model using all feature sets attained only a macro-F1 of 47\% ($SE=0.19\%$), indicating that discerning a user's gender pronoun category is more challenging. Textual features remain the best predictors of gender pronoun category at 46\% macro-F1 ($SE=0.19\%$), whereas activities features are the worst predictors at 35\% macro-F1 ($SE=2.50\%$). The feature set of top mentions is the best single set of features after the textual features, yielding a score of 42\% ($SE=1.12\%$).

\subsubsection{Highly Mentioned Users.} 
Since using the feature set of highly mentioned users alone achieves a non-negligible improvement over the baselines in both experiments, we further explore whether there is a link between gender pronoun disclosure and the gender pronoun categories of the 1K highly mentioned users. To this end, we first check if the 1K highly mentioned users belong to any gender pronoun group. 12\% of the highly mentioned users have gender pronouns (\textit{she} series: 6\%, \textit{he} series: 4\%, and non-binary: 2\%). We then compute the relative frequency of them getting mentioned by GP users as opposed to NP users. The result is depicted in Fig \ref{fig:rq2_mentions}. Across the board, all highly mentioned GP users are mentioned significantly more by GP users (Mann-Whitney U test, $p<0.001)$. These differences are also significant across all four groups of users using a Kruskal-Wallis H test ($p<0.001$). Highly mentioned users who use non-binary pronouns, in particular, are 10 times more likely to be mentioned by GP users than by NP users ($p<0.001$).

\section{Gender Pronoun Adoption}\label{sec:rq3}

Previously, we saw that the overall proportion of tweets with gender pronouns is rising (\S\ref{sec:rq1}) and that the highly mentioned GP users are more likely mentioned by GP users. In this section, we want to better understand the mechanisms underpinning gender pronoun adoption, with a focus on social network effects. 
\begin{figure}
    \centering
    \includegraphics[width=\linewidth]{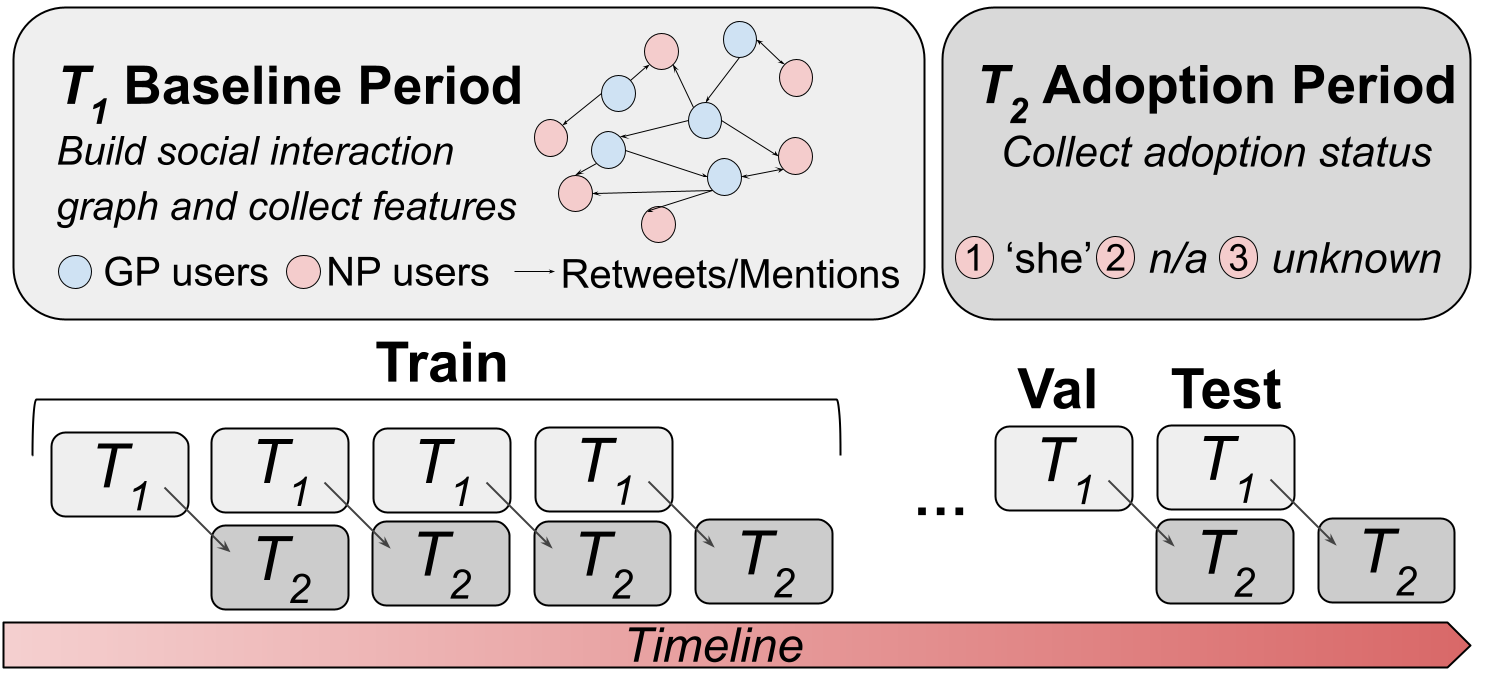}
    \caption{The workflow used to predict and analyze how prior Twitter usage and sharing activities inform future gender pronoun adoption.}
    \label{fig:adoption_workflow}
\end{figure}

\subsection{Method}

\subsubsection{Network Effects on Gender Pronouns.}
In any social network, either online or offline, ideas, emotions, and behaviors can diffuse from one to another, a process known as social contagion \cite{christakis2013social,ferrara2015measuring,kramer2014experimental,rosenquist2011social,tsvetkova2014social,pachucki2011social,hodas2014simple,monsted2017evidence,ferrara2015quantifying}. Prior works have also shown that people who congregate in social communities share similar traits, interests, and tendencies, resulting in homophilic communities \cite{mcpherson2001birds,kossinets2009origins,bisgin2012study}. Putting network homophily and contagion together results in the social network effect \cite{vanderweele2013social}. In this work, we elicit social network effects from observable sharing activities, such as retweeting and mentioning other users who disclose gender pronouns. While it is preferable to use the follower/following relationships among users \cite{hodas2014simple,ferrara2015measuring}, this is not computationally feasible in practice given Twitter API's rate limits. In lieu of the follower/following network, we use the retweet and mention networks under the assumption that the act of retweeting or mentioning can be treated as a form of a stronger, more explicit signal of social network exposure to peer influence as well as social endorsement \cite{boyd2010tweet,metaxas2015retweets}. 
\begin{table*}[]
    \centering
    % \footnotesize
    \small
    \begin{tabular}{ccrrrrr}
        \toprule
        \textbf{Network} &\textbf{Time Span}&  \multicolumn{2}{c}{\textbf{No. Users}} & \multicolumn{3}{c}{\textbf{Pronouns} (\% of Total)}\\
        \cmidrule(lr){3-4}\cmidrule(lr){5-7}
        &  (Wks)& Total &  Pronouns & \textit{She} Series & \textit{He} series & Non-binary  \\
         \midrule
         {Retweet} & Four &11,435,974 &165,590 (1.4\%) & 0.9\% & 0.3\% & 0.3\% \\
        {Retweet} & Eight  & 4,106,182 & 101,023 (2.5\%) & 1.6\% & 0.5\% & 0.4\% \\
         {Mention} & Four& 19,533,731  & 278,084 (1.4\%) &0.9\% & 0.3\% & 0.2\%  \\
        \bottomrule
    \end{tabular}
    \caption{The number of users included in the RQ3 experiments and the proportion of users who adopted gender pronouns in $T_2$.}
    \label{tab:rq3_data}
\end{table*}
\subsubsection{Defining Measurement Periods.} To measure the network effects on gender pronoun adoption, we restructure our data into temporally disjointed subsets. Consider two time windows of data $T_1$ and $T_2$, where $T_1$ occurred prior to $T_2$. We build a directed, weighted, and attributed network $G=(V,E)$ from tweets posted in $T_1$ as follows. We first identify the set of GP users $U_p$ who appeared in our dataset with gender pronouns during $T_1$. Then we extract the network interactions  $E=\{e\}$ that occurred in $T_1$, where each edge $e=(u,v, w)$ is a directed edge where at least one of the users involved in the interaction is a GP user, i.e., $u\in U_P$ or $v\in U_P$. The weight of the edge $w$ represents the frequency of the mention or retweet interactions from \textit{u} to \textit{v}.  In the context of this paper, retweeting a user is one way to mention them, therefore the retweet network edges are a strict subset of the mention network edges. We then retain the largest weakly connected component of the network. The NP users in the network compose the set $U_{NP}$, where $V=U_P\sqcup U_{NP}$. In the ensuing period $T_2$, we check if any of the users in $U_{NP}$ adopted gender pronouns. Note that a substantial amount of users would have unknowable pronoun status if we do not see their tweets in $T_2$. 

We repeatedly build $T_1$ and $T_2$ data using sliding windows of discretized time periods starting from Feb 1, 2020. We experiment with spans of four weeks for graphs built on both retweet edges and mention edges. This produces twenty-one $T_1$ and $T_2$ pairs, ending the last $T_2$ period on Oct 9, 2021. For graphs with retweet edges, we further explore a larger span of eight weeks to control for data biases due to time span selection. This produces ten $T_1$ and $T_2$ pairs, ending the last $T_2$ period on Oct 23, 2021. 

\subsubsection{Data Statistics.}
The adoption experiment data sets used in this section are described in Table \ref{tab:rq3_data}. We show the number of target users included in the experiments and how many of those eventually adopted pronouns. These users are all part of $U_{NP}$, that is, users who were connected to a GP user ($U_{P}$) in $T_1$ but who themselves did not have gender pronouns in $T_1$. 
Overall, the proportion of gender pronoun adoption is very low at around 2\%, rendering the data extremely imbalanced. This comes as no surprise given the relatively low gender pronoun usage we have seen so far in this paper, even though, by design, every user is connected to at least one GP user through retweeting or mentioning them. Of all the gender pronouns adopted, \textit{she} series is the most popular pick. \textit{He} series and non-binary pronoun adoptions follow closely behind, with \textit{he} series pronouns slightly more frequently adopted. 

\subsubsection{Features Sets.}
To explore the relationship between data collected from $T_1$ and the adoption of gender pronoun in $T_2$ of all users in $U_{NP}$, we use the following sets of features:
\begin{itemize}

\item \textbf{{Text}} ($\text{dim}=768$): the BERTweet embeddings of the tweets each user posted in $T_1$, concatenated as a single string.

\item \textbf{{User}} ($\text{dim}=768+6$): the BERTweet embedding of the user's biography and user-level metadata including friends count, followers count, favorites count, statuses count, listed count, and whether they are verified. We use the last entry in $T_1$ if a user appeared more than once.

\item \textbf{{Network}} ($\text{dim}=5+5+128$): the number of neighbors they have (i.e., they retweeted or mentioned), the number of GP neighbors, and the number of neighbors with \textit{she} series, \textit{he} series, or non-binary pronouns. We also compute all weighted versions of the previous numbers of neighbors. Additionally, we include the node2vec \cite{grover2016node2vec} graph embeddings learned from the retweet or mention network (see Appendix).

\item \textbf{{Activities}} ($\text{dim}=11$): the number of original tweets, re\-tweets, quoted tweets, or replies the user posted during $T_1$. We also count the number of hashtags, URLs, mentions, characters, and tokens used. Finally, we include the number of first-person singular pronouns used in the tweets, which convey individualism, and the number of first-person plural pronouns, which convey collectivism \cite{twenge2013changes}. 
\end{itemize}

\subsubsection{Modeling.} We conduct two experiments. Experiment (1) is the binary classification of whether the user will adopt gender pronouns in $T_2$ and experiment (2) is the multiclass classification of which of the three gender pronoun categories the user adopts in $T_2$. Users who did not adopt any gender pronouns are excluded from experiment (2), as was done in \S\ref{sec:rq2_modeling}. Our experiments are repeated for all data scenarios: retweet networks over four- and eight-week time spans and mention networks over four-week time spans (i.e., $T_i$ is either four or eight weeks). For each data scenario, we use the last $T_1$--$T_2$ pair as the test set, the second to last  $T_1$--$T_2$ pair as the validation set, and the remaining data as the training set. Similar to \S\ref{sec:rq2_modeling}, we built a deep neural network (DNN) for these tasks. We first tune the model hyperparameters via randomized grid search using the training set. Please see the Appendix for the exact model architecture and hyperparameter search space. Macro-F1 is used as the evaluation metric. Following \S\ref{sec:rq2_modeling}, we explore the results using each feature set in isolation as well as in combination, and we use the \texttt{Random} and \texttt{Majority} models as baselines.

\subsection{Results}

\subsubsection{Classification Performance.}
\begin{figure}
    \centering
    \includegraphics[width=\linewidth]{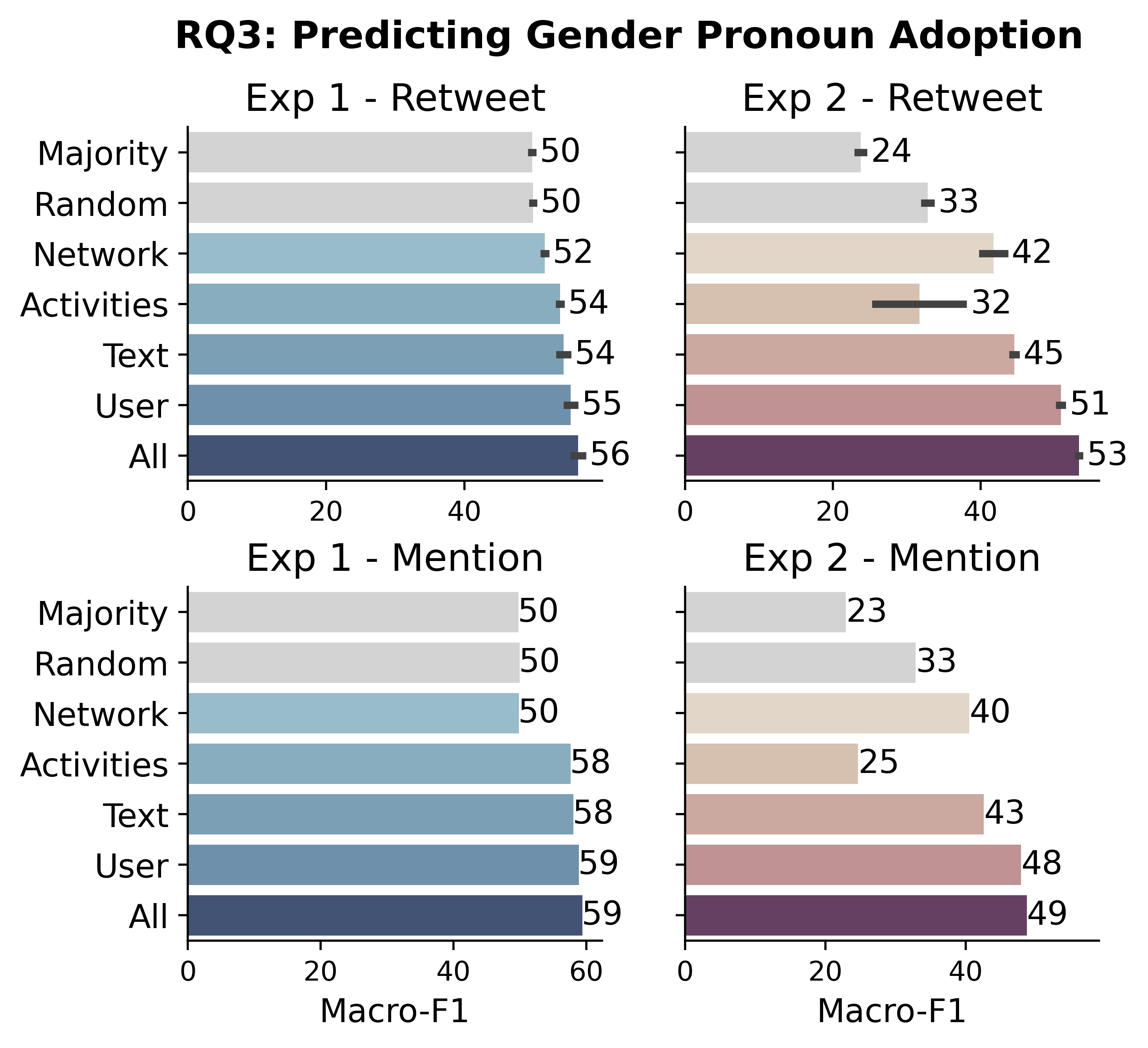}
    \caption{RQ3: Classification performance in the binary classification of detecting the presence of gender pronouns (Exp 1) and in the multiclass classification of the three gender pronoun types (Exp 2). The top row shows the results of the retweet networks, combining retweet networks spanning four weeks and eight weeks. The bottom row shows the results of the mention network spanning four weeks.}
    \label{fig:rq3_results}
\end{figure}

We present our results to address RQ3 in Fig. \ref{fig:rq3_results}. The experimental results of the four-week and eight-week time spans of the retweet network are averaged due to the similarity in their results. Differently from Section 5.2.1, we recognize lower classification performances in experiment (1), the binary classification of predicting whether a user will adopt any type of gender pronouns in $T_2$ given the features in $T_1$, probably due to the limited observation period leveraged in the training phase. Using all of the features sets in $T_1$ features, our approach performs slightly better than the \texttt{Random} and \texttt{Majority} baselines on the retweet network, achieving 56\% macro-F1 compared to the baseline score of 50\%. The mention network prediction tasks achieve better performance at scores of 59\% macro-F1. Combining all features yields very little improvement over using features in isolation. All singular categories of features perform relatively equivalently. User-level features are the most predictive, while network features are some of the lowest. We note that this could be an artifact of our study design limited by computational constraints, forsaking the larger social network of users who are not directly connected to a GP user. Therefore, we do not know how many other NP users they were connected to. The ineffectiveness of the network features becomes even more apparent using the mention network. One explanation is the underlying difference between retweets and mentions: retweets are often seen as endorsements \cite{boyd2010tweet,metaxas2015retweets} whereas mentions can represent social conversations \cite{leavitt2015influentials}. 

The predictive performance of experiment (2), which is the multiclass classification of which of the three gender pronoun category a GP adoptee chooses, yields further insights.  While user-level features remain the singularly most predictive set of features, we now see a large improvement gained from using network features. This shows that network features contain cues as to \textit{which} gender pronoun a user will adopt. We also see that the performance is better in general on the retweet network than on the mention network, suggesting that the retweet network signals encode more cues as to which gender pronoun category a user will adopt. This process could unfold via either social homophily or social contagion or both \cite{shalizi2011homophily}. Social networks are mediums for online communities of people with shared interests and similar backgrounds, forming homophilic clusters of users. It can also be a pathway for social influence, wherein one exerts behavioral changes on another by exposure. While we cannot disentangle the intertwining effects of the two forces, we provide additional analyses on the social network effects below.

\subsubsection{Relative Number of Pronoun Neighbors.}
To deepen our understanding of the link between the social network and gender pronoun adoption, we analyze how gender pronoun adoption in $T_2$ is linked to the relative number of GP neighbors they had in $T_1$. A user's neighbor is someone they retweeted in a retweet network and someone they mentioned in a mention network. We begin by establishing the expected number of type $x$-gendered neighbors any NP user would be connected to in $T_1$, given by:
\begin{equation}
    E[x] = \frac{1}{|U_{NP}|}\sum_{u\in{U_{NP}}}\mathcal{N}_u(x), 
\end{equation}
where $\mathcal N_u(x)$ denotes the number of $x$-gendered neighbors a user $u\in U_{NP}$ had in $T_1$, $x\in \{\textit{she}, \textit{he}, \textit{non-binary}\}$. The relative number of $x$-gendered neighbors for users who adopted $y$ gender pronouns in $T_2$ is thus given by:

\begin{equation}\label{eq:rel_neighbors}
R(x, y) = \left. 
\frac{1}{|U^{y}|}\sum_{u\in U^y}\mathcal{N}_u(x) 
\middle/ 
E[x],
\right.
\end{equation}
where $U^y\subset U_{NP}$ represents the set of users with pronoun status $y$ in $T_2$, $y \in\{\textit{she}, \textit{he}, \textit{non-binary},\allowbreak \textit{no pronouns}\}$. If $R(x, y) < 1$, then users who adopted pronoun status $y$ in $T_2$ had less $x$-gendered pronoun neighbors in $T_1$. Alternatively, if $R(x, y) > 1$, then users who adopted pronoun status $y$ in $T_2$ had more $x$-gendered pronoun neighbors in $T_1$. 

\begin{figure}
    \centering
    \includegraphics[width=\linewidth]{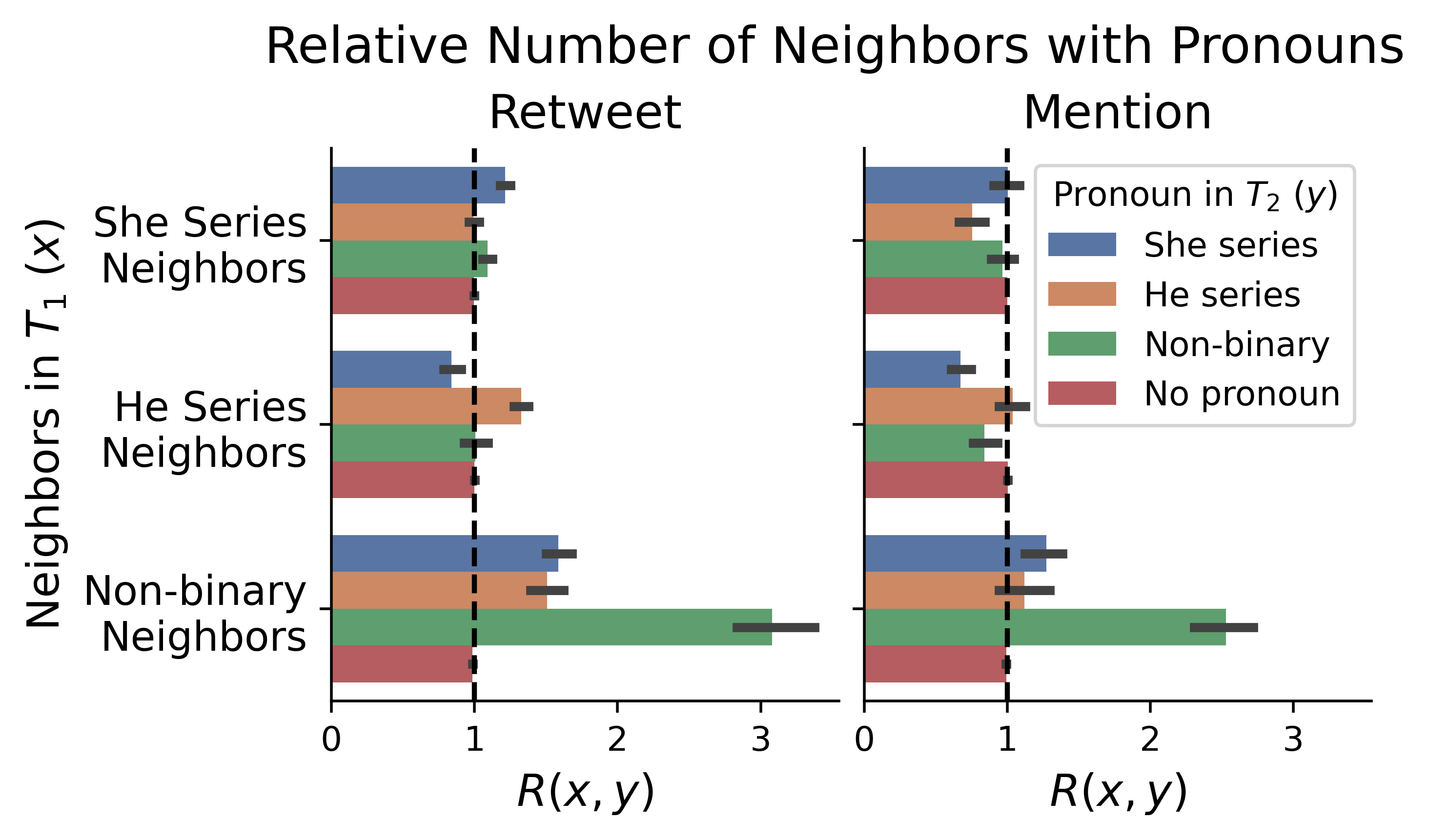}
    \caption{The relative number of pronoun neighbors a user has in the retweet and mention networks, given by Eq. \ref{eq:rel_neighbors}. Users who adopted non-binary pronouns in $T_2$ had an outsized number of non-binary neighbors in $T_1$.}
    \label{fig:rq3_rel_neighbors}
\end{figure}

Fig. \ref{fig:rq3_rel_neighbors} displays the relative number of neighbors $R(x,y)$ separately for the retweet and mention networks. We aggregate the statistics for each individual in every timeframe. The retweet networks of four-week and eight-week time spans are combined in this figure due to the similarity of their results. There are several key insights. First, we observe that users who did not adopt pronouns (red bars) retweeted or mentioned an expected number of GP users. That is, $R(x, \textit{no pronouns})\approx 1$. This is both a sanity check and a result that shows that users who remained without gender pronouns had network exposures that did not deviate from the norm. Interestingly, users who adopted \textit{she} and \textit{he} series pronouns retweeted and mentioned slightly more neighbors of the same gender pronoun category but slightly \textit{fewer} neighbors of the opposite gender pronoun category. 

Importantly, we see that users who adopted any gender pronouns \textit{at all} had substantially more neighbors who use non-binary pronouns. Those who adopted \textit{she} and \textit{he} series pronouns had 1.5 times non-binary neighbors in the retweet network. This number is particularly sizeable for non-binary pronoun adoptees, who retweeted over 3 times as many and mentioned over 2.5 times as many non-binary neighbors as the expected baseline.

\section{Discussion}
This paper provides a preliminary understanding of the landscape of gender pronoun usage on Twitter in recent years. We demonstrate the rising prevalence of gender pronoun usage and illustrate the differences between users with and without explicit gender pronoun disclosure, as well as between users who use \textit{she} series, \textit{he} series, and non-binary pronouns. We explain how social network influence leads to the adoption of gender pronouns. This work represents a critical initial step toward understanding the evolving landscape of gender pronoun use on social media. We discuss below the important findings and implications of this paper.

\subsubsection{RQ1: Gender Pronouns on Twitter.}

Our results show that gender pronoun usage is considerable on Twitter, with a rising number of tweets composed by users with gender pronouns over time. The most prevalent gender pronoun category is the \textit{she} series, followed by \textit{he} series and non-binary pronouns. \textit{She} series users overwhelmingly dominate \textit{he} series users at a ratio of 2:1, which could suggest that those who identify more as women are more prone to publicly disclose their gender pronouns than those who identify more as men. While there are fewer non-binary gender pronoun users, they nonetheless occupy a sizable proportion of the total Twitter population.

\subsubsection{{RQ2: Predicting Gender Pronouns.}}
We show that there are cues from a user's Twitter usage and sharing activities that can inform whether a user uses \textit{any} gender pronouns and also the type of gender pronoun they use. The implications are two-fold. On one hand, with textual data being the best predictor of gender pronoun status or lack thereof, there could be a difference in discourse revolving COVID-19 among men, women, non-binary people, and people who do not disclose gender pronouns. While previous Twitter work indeed showed that each gender identity group grapples with different COVID-19 concerns  \cite{al2021investigating}, we further provide evidence that discourse also differs between users with explicit gender pronoun disclosure and those who do not. On the other hand, non-textual features can also inform predictive modeling, suggesting that there are distinguishing patterns of social media usage that could help us discern users' proclivity to gender pronoun disclosure and users of various gender pronoun categories. We hope this work sheds light on future research directions on this matter.

\subsubsection{{RQ3: Gender Pronoun Adoption.}}
We show that gender pronoun adoption is linked to social network effects. There are two theories. The social contagion theory suggests that gender pronoun adoption is causally impacted by observing other users using gender pronouns \cite{christakis2013social}. The homophily theory suggests that users with similar propensities of disclosing gender pronouns naturally attract one another \cite{mcpherson2001birds}. Though it is virtually impossible to disentangle these theories in observational data \cite{shalizi2011homophily}, we believe that gender pronoun adoption is a much more complex process. Gender pronoun usage is acquiring increasing exposure at work \cite{wsj2021why}, at school \cite{waterloo}, and elsewhere in our lives \cite{dailymail2019identify,nbc2022word,nyt2022sports,policing2019trans}. It is therefore futile to decouple the effects of the world around us on the adoption of gender pronouns on Twitter.

With these precautions in mind, we make two conclusions regarding RQ3. The first is that we find different gender pronoun groups respond differently to social network stimuli, possibly indicating the users of each gender identity have distinct responses to influence. The second conclusion is that each category of gender pronoun users exerts varying levels of influence on gender pronoun adoption. Having non-binary pronoun users in the social network is linked to the adoption of \textit{any} gender pronouns, but is particularly linked to the adoption of non-binary pronouns. In other words, non-binary individuals stand out as being both the most likely to adopt gender pronouns after seeing others use gender pronouns and also the most related to others adopting gender pronouns in subsequent periods.

Finally, we must stress that our conclusions illustrate mechanisms of gender \textit{pronoun} adoption, not \textit{gender} or \textit{gender identity} adoption. Our results should be not misconstrued to imply that gender identity is infectious, but rather that there is a link between a user's social network interactions and their willingness to disclose gender pronouns. 

\subsection{Limitations}
There are several limitations of our work. First, we consider only public-facing Twitter users, which presents unavoidable self-selection and platform biases in the users included in this study. Second, we use a COVID-19 dataset, which offers a wide variety of users who may or may not use gender pronouns but may arguably bear biases. We note that a similar, parallel study on a random Twitter also reached similar conclusions \cite{tucker2022pronoun}.
% we treat all textual data, especially gender pronouns, as English, resulting in a Western-centric interpretation of our insights. 
Finally, our definition of social influence is self-selected exposure rather than involuntary exposure. Notwithstanding these limitations, we believe that our results carry actionable insights and crucial implications that could inform additional research in this area.

\subsection{Broader Impact}
Openly declaring one's gender nonconformity can be daunting \cite{inside2018problem}. Yet we witness an increasing number of people publicly sharing their non-binary identities. This paper underscores how observing gender pronoun usage can affect gender pronoun adoption. Even the active declaration of gender pronouns by cisgender people, however self-evident, is an expression of solidarity with those whose genders lie beyond the dichotomous sociocultural perceptions of gender. Thus, our work provides actionable insights toward improving gender inclusivity in online spaces by encouraging gender pronoun disclosure.

By focusing on more generalizable aspects of network dynamics and Twitter usage, our results also carry transferable implications to other situations, contexts, and mediums beyond strictly COVID-19 discussions. The omnipresence of the COVID-19 pandemic motivated us to use this dataset as a tool to study a heterogeneous sample of the Twitter population, including not only the marginalized LGBTQ+ community but also the wider cisgender society.

\subsection{Ethical Consideration}

We emphasize the ethical impact of our work. First and foremost, while we simplified the categories of gender pronouns to holistically analyze the landscape the gender pronoun usage on Twitter, we recognize the complexity and multiplicity of gender identities \cite{keyes2021you}. In this work, we also assume that a person's gender pronoun choice is a reflection of their gender identity, which may not be the case in reality. Furthermore, since we demonstrate the possibility to detect genders on social media, we are aware of how our research could be misused and abused, endangering the LGBTQ+ community \cite{unstruggle,hrc2020epidemic,cnn2020florida,gyamerah2021experiences,hrw2015outlawed}. As such, we avoid exposing harmful personal details and our analyses are carried out only at an aggregated, high level. We will also not share the raw data or our pronoun labels, which could be exploited to identify individual users. We also strongly encourage future studies to consider the ethical aspects of online gender-related studies from the initial study design to the final research output. 

\textit{Note: This study and the data used are IRB-approved.}

\section{Acknowledgments}

The authors are indebted and grateful to people from the LGBTQ+ community who offered their feedback on this work, especially Os Keyes (University of Washington) for their generous and helpful advice. We also thank Prof. Jonathan May (USC) for his helpful insights. This work was supported by DARPA (award number HR001121C0169).

{
\fontsize{9pt}{10pt} \selectfont
\bibliography{main}

}

\appendix
\section{Appendix}
\subsection{All Pronoun Keywords} We use the following list of keywords to match pronouns used in the texts for the RQ2 experiments conducted in \S\ref{sec:rq2}:
\begin{itemize}
    \item \textbf{I}: I, me, mine, my, myself
    \item \textbf{You}: you, your, yours yourself, yourselves 
    \item \textbf{We}: we, us, our, ours, ourself, ourselves
    \item \textbf{He}: he him his himself, 
    \item \textbf{She}: she, her, hers, herself 
    \item \textbf{It}: it, its, itself 
    \item \textbf{They}: they, them, their, theirs, xe, xem, ze, zem
\end{itemize}

\subsection{Node2vec}
We apply the node2vec algorithm \cite{grover2016node2vec} on the networks generated for the RQ3 experiments conducted in \S\ref{sec:rq3}, treating networks as weighted and directed. We use the default parameters, which fixes the dimension size to 128, the length of the walk per source to 80, the number of walks per source to 10, the context size to 10, the return hyperparameter to 1, the in-out hyperparameter to 1, and the number of epochs to 1. 
\subsection{Deep Neural Network Architecture}
The model architecture includes an input normalization layer, followed by a variable number of hidden layers. Each hidden layer is preceded by a dropout layer and succeeded by a normalization layer. The activation functions are ReLU. We use the Adam optimizer. The data is trained with batch sizes of 4096. All models are trained for 100 epochs with early stopping imposed if the validation loss stops reducing.

The binary classification models use the sigmoid activation function as the final layer activation and are optimized via binary cross-entropy loss. The multiclass classification models use the softmax activation as the final layer activation and are optimized via categorical cross-entropy loss.

\subsection{Hyperparameter Tuning}
For every prediction task, we conduct hyperparameter tuning using 40 randomized combinations of the following hyperparameters:
\begin{itemize}
    \item Number of layers: \{2, 4, 16\}
    \item Hidden units: \{32, 128, 768\}
    \item L2 regularization: \{$1\times 10^{-2}$, $1\times 10^{-3}$, $1\times 10^{-5}$\}
    \item Learning rates: \{$1\times 10^{-3}$, $1\times 10^{-5}$, $1\times 10^{-7}$\}
    \item Dropout: \{0.1, 0.2, 0.3\}
\end{itemize}

\end{document}